\documentclass[%
 reprint,
superscriptaddress,
 amsmath,amssymb,
 aps,
prx,
]{revtex4-1}

\usepackage{graphicx}
\usepackage[dvipsnames]{xcolor}
\usepackage{amssymb}
\usepackage{amsmath}
\usepackage{amsbsy}
\usepackage{color}
\usepackage{float}
\usepackage{bm}

\def\vec#1{\mbox{\boldmath $#1$}}

\usepackage{bm}

\usepackage{siunitx}

\begin{document}

\title{Organisation and dynamics of individual DNA circles in Kinetoplast DNA}

\title{Organisation and dynamics of individual DNA segments\\ in topologically complex genomes}

\author{Saminathan Ramakrishnan}
\affiliation{School of Physics and Astronomy, University of Edinburgh}

\author{Auro Varat Patnaik}
\affiliation{School of Physics and Astronomy, University of Edinburgh}

\author{Guglielmo Grillo}
\affiliation{Physics Department, University of Trento, via Sommarive, 14 I-38123 Trento, Italy}
\affiliation{INFN-TIFPA, Trento Institute for Fundamental Physics and Applications, I-38123 Trento, Italy}

\author{Luca Tubiana}
\affiliation{Physics Department, University of Trento, via Sommarive, 14 I-38123 Trento, Italy}
\affiliation{INFN-TIFPA, Trento Institute for Fundamental Physics and Applications, I-38123 Trento, Italy}

\author{Davide Michieletto}
\affiliation{School of Physics and Astronomy, University of Edinburgh}
\affiliation{MRC Human Genetics Unit, Institute of Genetics and Cancer, University of Edinburgh}
\thanks{davide.michieletto@ed.ac.uk}

\newcommand{\dmi}[1]{\textcolor{RoyalBlue}{#1}}

\begin{abstract}
Capturing the physical organisation and dynamics of genomic regions is one of the major open challenges in biology. The kinetoplast DNA (kDNA) is a topologically complex genome, made by thousands of DNA (mini and maxi) circles interlinked into a two-dimensional Olympic network. The organisation and dynamics of these DNA circles are poorly understood. In this paper, we show that dCas9 linked to Quantum Dots can efficiently label different classes of DNA minicircles in kDNA. We use this method to study the distribution and dynamics of different classes of DNA minicircles within the network. We discover that maxicircles display a preference to localise at the periphery of the network and that they undergo subdiffusive dynamics. From the latter, we can also quantify the effective network stiffness, confirming previous indirect estimations via AFM. Our method could be used more generally, to quantify the location, dynamics and material properties of genomic regions in other complex genomes, such as that of bacteria, and to study their behaviour in the presence of DNA-binding proteins. \\
\end{abstract}

\maketitle

\section{Introduction}
Understanding the spatial organization of complex and large genomes is currently one of the biggest challenges in biology and biophysics~\cite{Dekker2013,Hildebrand2020}. The kinetoplast DNA (kDNA), the mitochondrial genome of parasites of the class \emph{Kinetoplastida}, such as trypanosomes, is a large ($\sim$ 10-100 Mbp) complex genome made of interlinked DNA circles. Historically, trypanosomes and the kDNA more specifically, have been at the centre of active research due to its role in pan-genomic RNA editing~\cite{Simpson1967,Simpson2000,Hajduk2010}. More recently, the kDNA has also been studied by the polymer physics and topology community as it is the archetype of a so-called ``Olympic network'', i.e. a structure formed by thousands of topologically concatenated rings~\cite{Diao2004,Michieletto2014,Klotz2020,Tubiana2024,Yadav2023,Soh2020,Soh2021,He2023,Ramakrishnan2024,Michieletto2025}. Such structures are rare because challenging to controllably synthesise in the lab~\cite{Panoukidou2024,Grosberg2020,Speed2024}.

There are many open questions on the self-assembly, replication and structure of kDNAs~\cite{Perez-Morga1993,Liu2005,Klingbeil2001a,Hoffmann2018,Kalichava2021,Lukes2002,Amodeo2021,Amodeo2022}.
For instance, \emph{Crithidia fasciculata} (\emph{C. fasciculata}) kDNA is made of around 5000 short minicircles (2.5 kbp) that are split in 18 genetic classes~\cite{Ramakrishnan2024}, and of 30 longer maxicircles, around 30 kbp each. While the biological role of these classes of DNA circles in encoding for messenger and editing guide RNAs is known, there are no quantitative measurements of the spatial location of individual genetic classes and whether they are segregated within the network or uniformly distributed~\cite{Lukes2005}. The mechanisms of kDNA replication and reorganization of maxi and minicircles are intriguing: during replication, minicircles are decatenated, replicated, and reattached to the periphery, while maxicircles do not decatenate from the network but undergo rolling circle amplification and redistribute inside the kDNA with the help of the enzymes~\cite{kinetoplastreplication,Kinetoplastreplication2,nabelshnur}. The ``nabelschnur'' structure~\cite{nabelshnur}, likely formed by maxicircles, is found to be crucial for the faithful segregation of replicated kDNA in daughter cells. Thus, kDNA maxicircles may have significant structural role in achieving the elongation and segregation of the kDNA~\cite{nabelshnur}. However, the organization of maxicircles within the network remains unknown and challenging to quantify.  

\begin{figure*}[t!]
    \centering
    \includegraphics[width=1.0\textwidth]{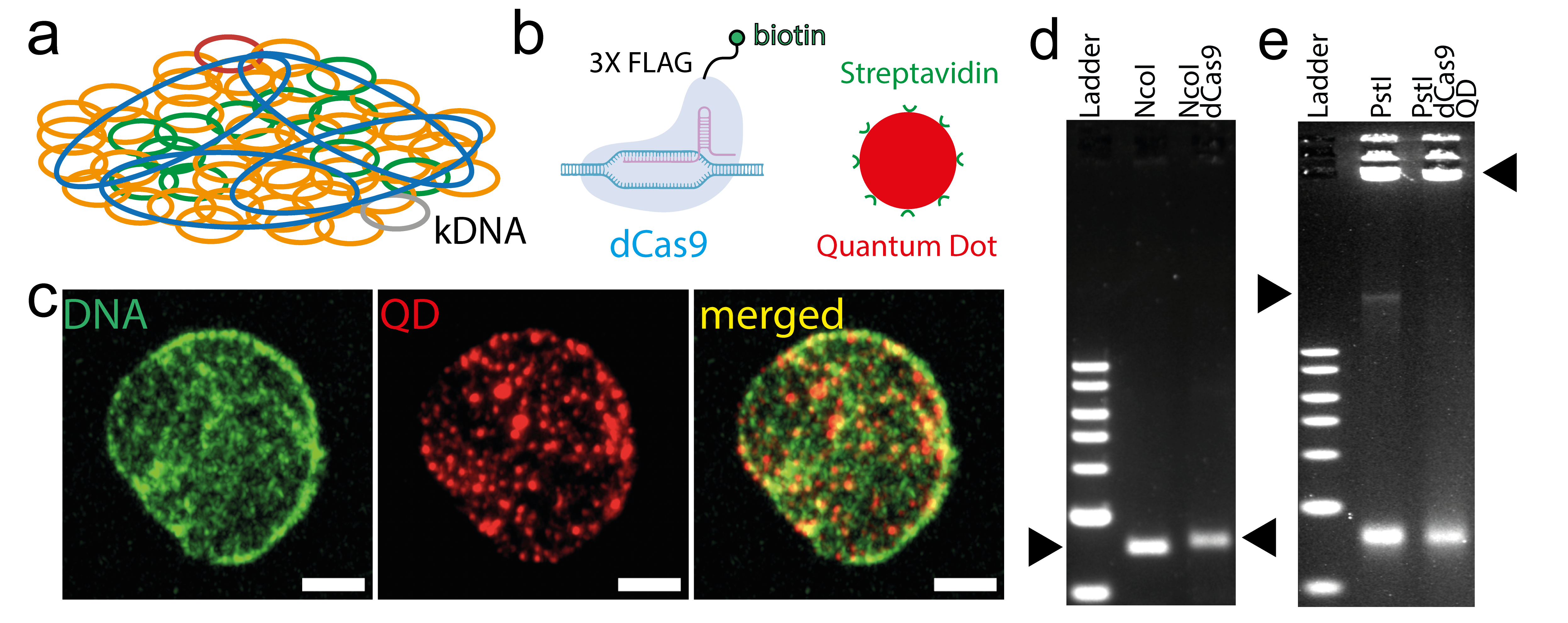}
    \vspace{-0.8 cm}
     \caption{\textbf{dCas9-labelling of kDNA sequences \textit{in vitro}.} \textbf{a} Sketch of \textit{C. fasciculata} kDNA showing major class minicircles in orange, minor class in green and maxicircles in blue. \textbf{b} To visualise the different classes of DNA circles we employ dCas9 and streptavidin-coated Quantum Dots (Qdot655). \textbf{c} Representative image of a kDNA structure with major minicircle class labelled by Qdot655. \textbf{d} Agarose gel electrophoresis of control and dCas9-bound minicircles, both cleaved with NcoI. The dCas9-bound minicircles shift upward in the gel because heavier than the control. \textbf{e} Gel of kDNA maxicircles cleaved with PstI (left arrow). When dCas9 proteins are targeting the maxicircles and are bound by Qdot655, the maxicircles form large complexes that remain stuck in the wells with the uncleaved part of the kDNA.} 
     \vspace{-0.4 cm}
    \label{fig:labelling}
\end{figure*}

In this paper, we employ a catalytically dead Cas9 (dCas9) proteins tagged using quantum dots (QDs) as physical beacons to identify target DNA sequences in \emph{C. fasciculata} kDNA and to quantify their spatial location and dynamics within the network \textit{in vitro} (Fig.~\ref{fig:labelling}a-c). We discover that both maxicircles and the major class of minicircles are enriched at the periphery, while the minor class of minicircles are uniformly dispersed within the network. Through computer simulations, we provide evidence that the location of maxicircles at the periphery and their linking to the minicircles induces the buckling of kDNA in solution, as seen in experiments~\cite{Klotz2020,Yadav2023}. Additionally, we are able to track the dynamics of individual sequences within kDNA maxicircles and their relative displacements. We discover that the dynamics of the maxicircles within the kDNA display a subdiffusive, correlated behaviour that reflects the high level of entanglement. Additionally, we use these measurements to estimate, for the first time, the effective stiffness of the kDNA network which confirms previous indirect estimations from AFM images~\cite{He2023}.

We argue that our method could be used more broadly to quantify the static and dynamic behaviour of specific, individual genomic regions within complex genomes, also in presence of DNA binding proteins such as transcription factors, and ultimately inform their material properties and dynamics.

\section{Methods}

\subsection*{Selection and preparation of sgRNA}

CRISPR RNA (crRNA) target sequences in maxicircles and minicircles were identified using the ChopChop online tool (\href{https://chopchop.cbu.uib.no/}{https://chopchop.cbu.uib.no/}) using the sequences we obtained in Ref.~\cite{Ramakrishnan2024} and purchased from Integrated DNA Technologies (IDT). The sgRNA was assembled with crRNA and universal trans-activating CRISPR RNA (tracrRNA) following the instructions provided by IDT. To assemble the sgRNA structure, 10 $\mu$M of crRNA and 10 $\mu$M of tracrRNA were mixed in IDT duplex buffer. The mixture was heated to 95$^\circ$C for 5 seconds and then allowed to cool down on ice for 1 hour. The sgRNA samples were aliquoted and stored at –20$^\circ$C.  The crRNA targets for kDNA maxicircles (24398 bp) are AGAGGCATCGAAGGATTGAGGGG (seq:5393), TTGAACGAGAATCCTGTATGCGG (seq:23384), AGGTACAACACCATAACACAGGG (seq:13854). The targets for the major minicircle class (2525 base pairs) is GGGCCGAGTGTTCTTGCACGAGG (seq:1678), while for the minor minicircle class (2538 base pairs) is CCGTCGGCAGAAATAGACCTGGG (seq:1727).
 

To validate the binding of dCas9 to kDNA maxicircles and minicircles, 400 nM of dCas9 was incubated with 800 nM of specific sgRNA (targeting three sites in the maxicircle and one site in both major and minor minicircle classes) in 1X NEB r3.1 buffer at 27$^\circ$C for 30 minutes. We then mixed the sgRNA-dCas9 complex with 500 ng of kDNA and incubated at 37$^\circ$C for 4 hours. The samples were digested with NcoI (for major class minicircles) and with BamHI (for minor class minicircles) at 37$^\circ$C for 60 minutes (see Results section). 


For the maxicircles, the binding of dCas9 would not yield a clear shift due to the size of the maxicircles and gel resolution. For this reason, we prepared the binding validation assay as follows: 500 ng of kDNA was mixed with sgRNA-dCas9 complex in 1X NEB r3.1 buffer. The mixture was incubated at 37$^\circ$C for 4 hours and then the sample was purified by agarose gel filtration and digested with 1 $\mu$L of PstI at 37$^\circ$C for 60 minutes. Finally, before gel analysis, the maxicircle-digested sample was incubated with 10 nM of Qdot655 Streptavidin Conjugate (Thermo Fisher) at 37$^\circ$C for 5 minutes, and then analyzed by 1\% agarose gel (see Results section). 

\subsection*{Microscopy of dCas9-kDNA-Qdot655 samples}
To prepare a dCas9-kDNA-Qdot 655 complex, 400nM dCas9 was first mixed with 800nM of sgRNA (3 targets) in 1X NEB r3.1 buffer and incubated at 27$^\circ$C for 30 minutes. The dCas9-RNA samples were pooled together and mixed with 500 ng of kDNA in 1X NEB r3.1 buffer and incubated at 37$^\circ$C for 4 hours. All samples were kept on ice after the respective incubation periods prior to the microscopical analysis. The kDNA-dCas9 sample was loaded onto a 1.5\% agarose gel and run at 80V for 20 minutes and recovered in 1x NEB r3.1 as described in detail in our previous paper Ref.~\cite{Ramakrishnan2024}. To prepare the samples for fluorescence microscopy, a clean glass coverslip treated with Poly-lysine was used. A 10 $\mu$L of recovered kDNA-dCas9 complex in NEB r3.1 buffer was mixed with 1 $\mu$L of 10 nM Qdot655 Streptavidin Conjugate and incubated at RT for 5 mins. A 5 $\mu$L sample was aliquoted into a separate tube and mixed with 1 $\mu$L of 10 nM YOYO-1. The mixture was then placed on a clean glass coverslip and sealed for fluorescence microscopy imaging.

The samples were imaged with a Zeiss LSM980 Airyscan2 laser scanning confocal microscope (Zeiss UK, Cambridge). A 63x/1.4 NA oil immersion objective was used, with 488 nm/633 nm excitation lasers for YOYO-1 and Qdo 655 respectively. Single time-point, 32 slice Z-stacks of the individual kDNA structures with dCas9-Qdot 655 were recorded using the Airyscan detector with a pixel size of 35 x 35 x 160 nm. Images were then Airyscan processed using Zen blue 3.5 software (Zeiss).  

To capture the real-time videos for dynamics analysis, 1 $\mu$l of YOYO-1-stained Qdot655-kDNA sample was suspended in 4 $\mu$L of 70\% glycerol, pipetted onto a glass coverslip, and sealed with a sticky spacer. Before acquisition, single suspended kDNA structures were carefully focused on the YOYO-1 channel and selected based on their visible outer ring structure. For each condition at least 25 movies were captured. Movies were recorded in confocal mode exciting with the 405 nm and 488 nm laser simultaneously exciting and capturing YOYO-1 (detector wavelengths 491-610 nm) and Qdot 655 (detector wavelengths 658-755 nm) at 8 fps, with a pixel size of 70 x 70 nm and at least 500 frames per video. Images were later deconvolved using Hugyens Professional 23.10 software (SVI). The fluorescence microscopy images presented in the manuscript were processed in FIJI (National Institutes of Health) and custom written python codes (see below). 

Example images obtained with this method are shown in Fig.~\ref{fig:distrmaxi}a-b, where the dCas9 is targeting maxicircle sequences in control (Fig.~\ref{fig:distrmaxi}a) or PstI-treated (Fig.~\ref{fig:distrmaxi}b) kDNA samples. In the latter, since PstI cleaves all maxicircles and the samples are gel purified, we expect and indeed observe no Qdot signal above the noise.

\subsection*{Image analysis}

The kDNA images were first processed using Gaussian smoothing to enhance the contrast between the sample and the background. After smoothing, a thresholding was applied to isolate the kDNA. Subsequently, OpenCV’s Canny Edge Detection was employed to extract the boundary pixels of the kDNA.
With this method we could robustly exclude any Qdot 655 located outside the kDNA boundary and measure the Euclidean 2D distance of each QD from the nearest kDNA boundary pixel, without the need to assume a simple circular shape.

\begin{figure*}[t!]
    \centering
    \includegraphics[width=1.0\textwidth]{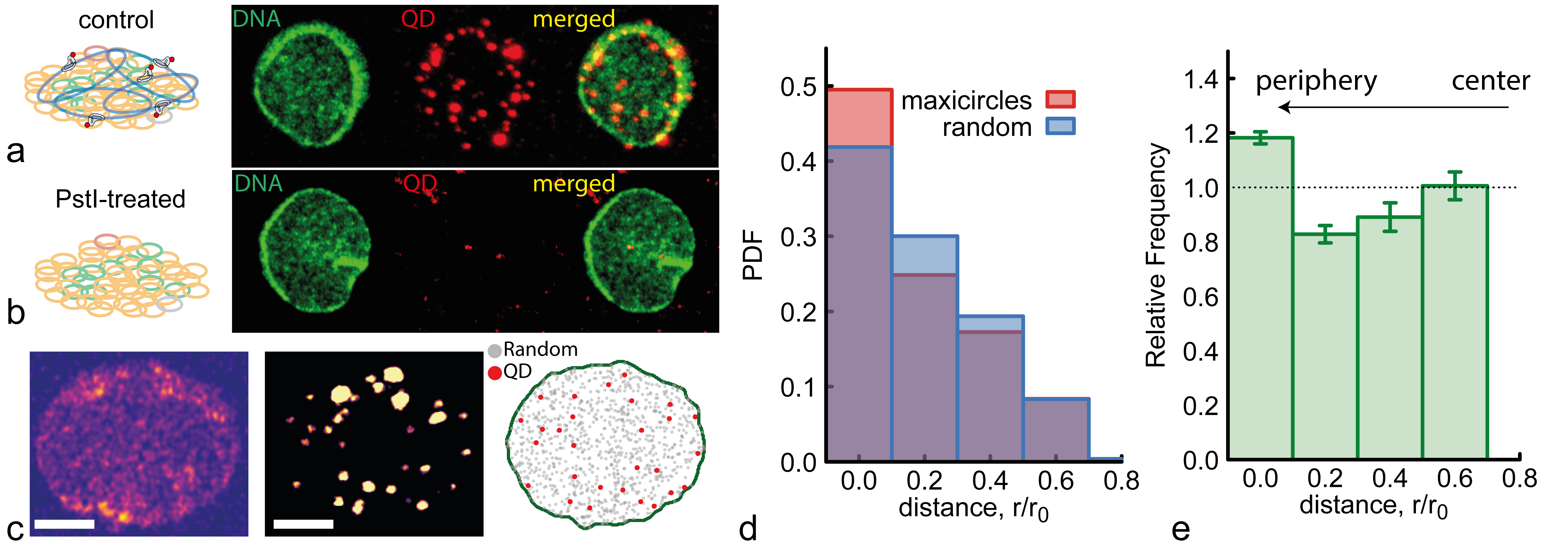}
    \vspace{-0.8 cm}
     \caption{\textbf{Maxicircles are preferentially located at the periphery.} \textbf{a-b} Representative images of YOYO-I labelled kDNA (green) and QDs (red). \textbf{c} Image analysis pipeline to quantify the distribution of Qdot 655   within the kDNA: images are Gaussian blurred and thresholded to detect the boundary of the network and the QDs. The Euclidean 2D distance of each QD from the closest boundary pixel is measured. The same calculation is performed on a simulated kDNA boundary with random points (1000 times the number of Qdots found in the respective kDNA) to obtain a random distribution of distances for individual kDNAs. \textbf{d} Histograms of distances of Qdot 655   (red) and random (blue) points as a function of distance from the kDNA boundary. \textbf{e} Normalised relative frequency (observed/random) as a function of distance from the kDNA boundary. One can appreciate a significant enrichment of localisations close to the boundary. } 
     \vspace{-0.4 cm}
    \label{fig:distrmaxi}
\end{figure*}

To quantify the specific location of the Qdots bound to the different kDNA sequences, we compared the distribution of Qdots with a distribution of points (100 times more than the count of Qdots) obtained by randomly sampling the space within the boundary pixels. The observed Qdot655 count per bin was then normalized by this simulated random distribution to reveal any specificity and biases in the location of the sequences (see Fig.~\ref{fig:distrmaxi}c).

Tracking of the QD was done from images taken using a spinning-disk confocal microscope. We then used trackpy (github.com/soft-matter/trackpy) on the QD signal to reconstruct their XY movement. To track and the kDNA centre of mass we used an in-house Mathematica code to perform a Gaussian blur and segmentation on the DNA signal, and then to obtain the centroid of the kDNA so to track its XY position during the experiment. 

\subsection*{Molecular Dynamics Simulations}
We performed molecular dynamic (MD) simulations of three different network type. The first system is a network of $604$ semi-flexible minicircles catenated with a valency of $3$, labelled ``minicircle only'' (MO). The second has three interlinked maxicircles linked intertwined throughout the minicircle disk, labelled ``linked diffuse'' (LD). The third has three interlinked maxicircles linked to the border of the minicircle disk, labelled ``linked border'' (LB). In all the configurations, each minicircle is composed of $m_{mini}=60$ beads while the three maxicircles have around $m_{maxi}=800$ beads each. Both circle species have a persistence length of $l_p=4\sigma$ and FENE bonds between beads. The minicircle networks are built using NetworkX~\cite{SciPyProceedings_11}, following the same procedure adopted in ref.~\cite{He2023}. LD and LB networks are obtained by slightly compressing the miniring networks inside a slit and then randomly intertwining the maxicircles with the minicircles network. The initial compression of the MO network is intended to reproduce the in-vivo conformation of the minirings, a condensed disk-like shape, without relying on further assumptions on its molecular origin. The detail procedure to construct the networks is reported in the SI. 

All the systems are evolved using an underdamped Langevin dynamics $\gamma = 0.1 $ and time step $dt = 0.01 \tau_{LJ}$, where $\tau_{LJ}$ is the characteristic time of the simulation. Equilibration is run for $10^8$ timesteps and production is run for at least $1.5\times 10^9 $ timesteps. We analyzed the curvature of the simulated kDNAs by using libIGL for Python and creating a triangulated surface from the COM of the minirings (see ref.~\cite{He2023}).

\section{Results}

\subsection*{dCas9 can specifically bind to both mini and maxi circles}

First, we investigated the binding of dCas9 to mini and maxicircles using Electrophoretic Mobility Shift Analysis (EMSA). We selected three target sites within the kDNA maxicircles, each spaced 5,000 base pairs apart, along with a single target in each minicircle class~\cite{Ramakrishnan2024}. To assess dCas9 binding to specific DNA types, we first allowed dCas9 to bind to its target sites, followed by digestion of the DNA using sequence-specific enzymes. In both the major and minor minicircle classes, a clear shift in the dCas9-bound DNA bands was observed, confirming sequence-specific, single dCas9 binding (Fig.~\ref{fig:labelling}d-e). To distinguish dCas9-bound maxicircles from the control, the complex was further incubated with streptavidin-coated Qdot655, which binds to biotin-labelled dCas9. The disappearance of the maxicircle band in the gel (Fig.~\ref{fig:labelling}e) confirmed the specific binding of Quantum Dots to the maxicircles. Interestingly, the presence of dCas9 slightly reduced DNA digestion compared to the protein-free control samples, although this did not affect the outcome of the EMSA assay. This effect is possibly due to the reduced 1D diffusion of restriction enzymes on DNA (see SI). Thus, we confirm that dCas9 can specifically bind mini and maxicircles sequences within the kDNA structure. 

\begin{figure*}[t!]
    \centering
\includegraphics[width=0.95\textwidth]{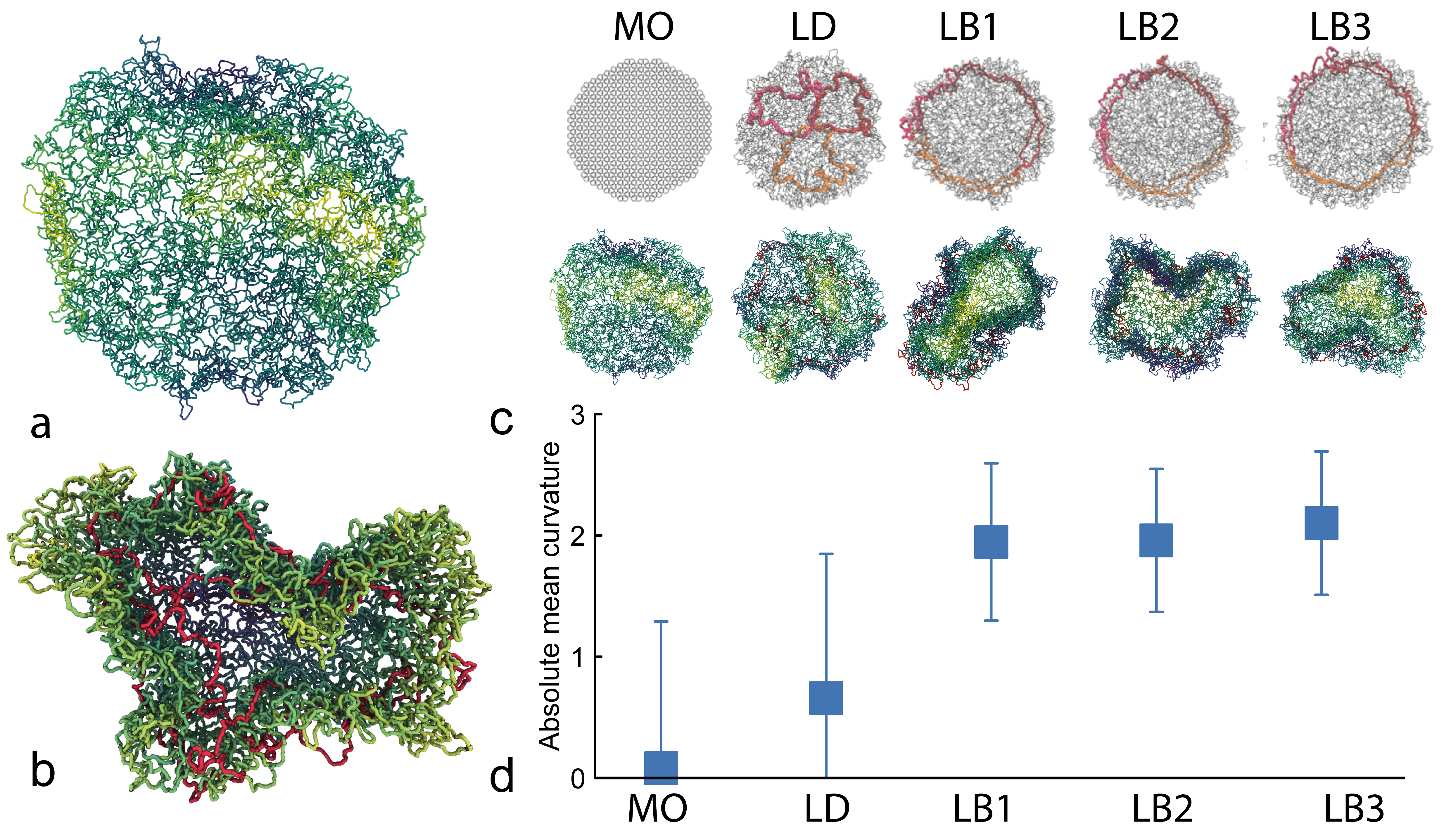}
     \caption{\textbf{MD simulations of kDNA suggest peripheral maxicircles induce buckling.} \textbf{a-b.} Representative snapshots of equilibrium configurations of kDNA networks without (a) and with (b) maxicircles. \textbf{c.} Snapshots of initial and equilibrium conformations of different kDNA topologies. (MO = minicircle only, LD = linked diffuse, LB1,2,3 = linked border 3 different replicas). \textbf{d.} Absolute mean curvature for the different kDNA topologies. The LB models display the largest mean curvature, reflective of buckling induced by the maxicircles linked at the border.} 
     \vspace{-0.4 cm}
    \label{fig:simbuckling}
\end{figure*}

\subsection*{Maxicircles are preferentially located at the periphery of the kDNA}

Having confirmed the specific binding of dCas9 to different kDNA circles, we then investigated the spatial distribution of different DNA sequences within the network. To do this, we first targeted three specific maxicircle sequences for dCas9 binding and then incubated the kDNA-dCas9 complex with QDs. Before imaging, we also removed the excess unbound dCas9-QD by gel filtering (as in Ref.~\cite{Ramakrishnan2024}). The resulting purified kDNA-dCas9-QD complexes were visualized using confocal fluorescence microscopy. Images were captured separately in the green YOYO-1 channel (labelling the kDNA) and the red 655 channel (labelling the QDs), and later reconstructed into a single composite image to reveal the spatial organization of maxicircles (Fig.~\ref{fig:distrmaxi}a-b). Bright fluorescent QD signal was visible at the periphery of the kDNA, with some signal appearing larger, suggesting potential clustering or co-localisation of dCas9-QD. Additionally, we detected located within the central region of the kDNA. To quantify the distribution of QDs, we segmented the DNA signal and reconstructed its boundary. We then segmented and localised the QD signal and placed it within the reconstructed kDNA boundary (Fig.~\ref{fig:distrmaxi}c). To compare the distribution of QDs with respect to a uniform distribution, we generated a random deposition process of 100 times more QDs within the same kDNA area. We then computed, for each QD (either randomly placed or real), its distance from the closest boundary pixel, and binned these distances to obtain distributions. The pipeline is represented in Fig.~\ref{fig:distrmaxi}c where random localisations are shown in grey and real QD localisations shown in red, all placed within the kDNA boundary (green). The analysis pipeline can be found open access at \url{https://git.ecdf.ed.ac.uk/taplab/kdnart}. 

In Fig.~\ref{fig:distrmaxi}d we compare the probability density function (PDF) of the random dots (blue boxes) with the one from the real QDs (red boxes). The relative distance, $r/r_{0}$, represents the binned distance of a localisation from the closest kDNA boundary pixel, normalized by the maximum radial distance of boundary pixels from the centre of the kDNA. In other words, an $r/r_{0}$ value of $0$ indicates that the localisation is at the periphery, while a value of $1$ corresponds to a localisation in the middle of the kDNA. The normalization allows us to sum the data across 13 individual kDNA networks, which have slight variations in their diameter. 

To identify any specific enrichment, we then divide (bin by bin) the observed distribution probability (PDF) of the QDs with that of the random, simulated localisation for each kDNA sample. A relative frequency of $1$ in a given bin indicates that we statistically find as many QDs as expected for a random (uniform) distribution (see Fig.~\ref{fig:distrmaxi}d). From this relative frequency one can appreciate that the distribution of QD is not uniform as a function of distance from the periphery. Interestingly, there is a significant enhancement within the 20\% of kDNA area closest to the periphery and lower than random in the kDNA area closest to the centre. This suggests that the maxicircles have a preference for localizing to the periphery of the network. Also, no maxicircle sequence is found within the 20\% area closest to the kDNA centre.

\begin{figure*}[t!]
    \centering
    \includegraphics[width=0.95\textwidth]{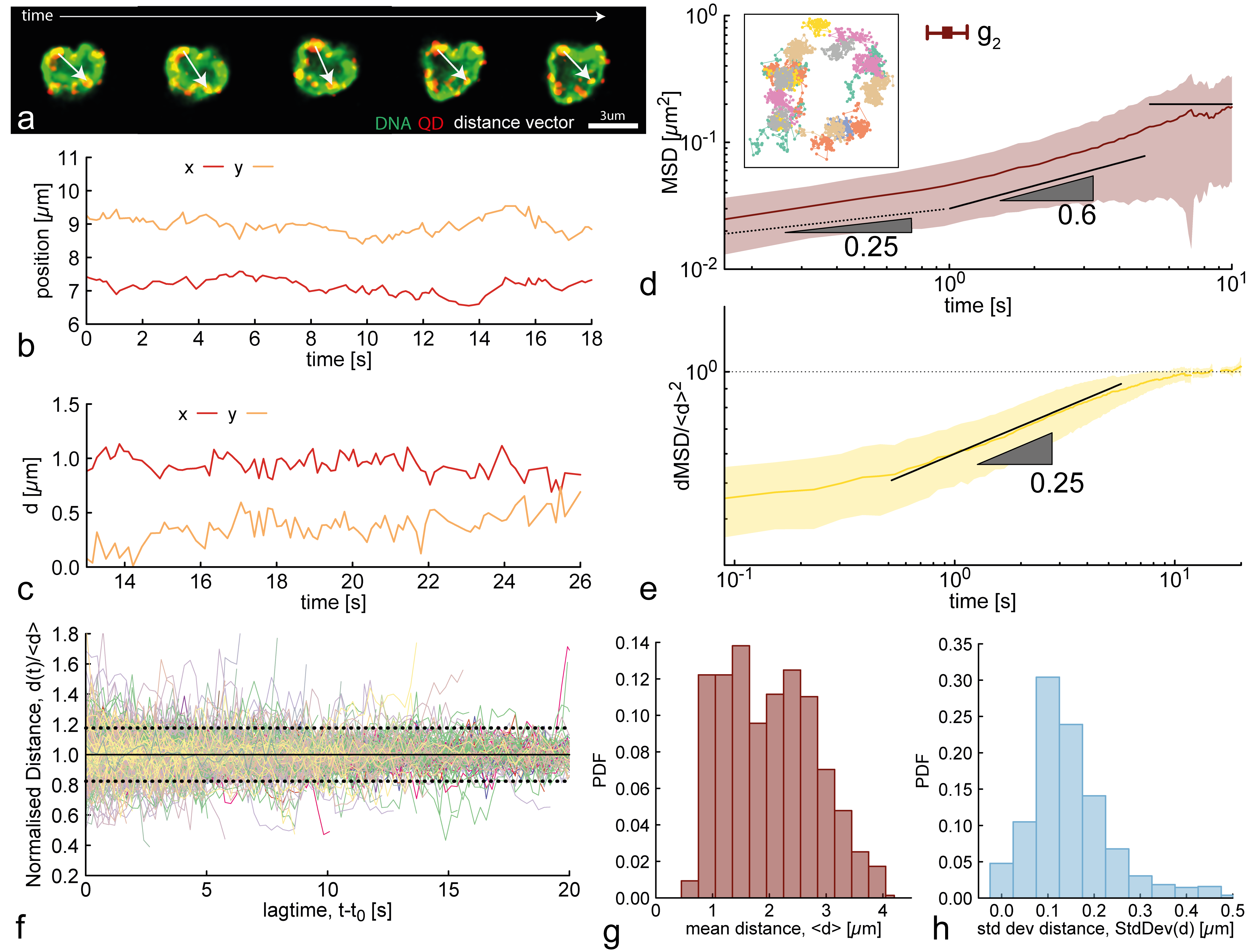}
     \caption{\textbf{Dynamics of maxicircle sequences.} \textbf{a.} Five snapshots taken with a spinning-disk confocal microscope of a fluctuating kDNA in solution at sequential times. DNA signal in green, QD signal in red. The white arrow indicates the distance vector between two QDs. \textbf{b.} Example of the position of one QD (x and y coordinates) over time. \textbf{c.} Example of the distance vector between two QD (x and y components) over time. \textbf{d.} Mean squared displacement (MSD) of the QDs in the frame of reference of the kDNA COM ($g_2(t)$).  \textbf{e.} Mean squared displacement of the distance vector $\vec{d}$ between pairs of QDs and normalised by the mean distance (squared) of each pair, $\textrm{dMSD}/\langle d \rangle^2$, averaged across pairs and initial times. \textbf{f} Normalised distance between pairs of QDs. Dashed lines represent one standard deviation from the mean. \textbf{g} Distribution of mean distances between QDs. \textbf{h} Distribution of standard deviations of distances between QDs. } 
    \label{fig:dynamics}
\end{figure*}

\subsection*{A peripheral distribution of linked maxicircles explains kDNA buckling in solution}

In solution, kDNA assumes a buckled shape with positive overall mean curvature that resembles a showercap~\cite{Klotz2020,Yadav2023,Soh2020}. This is surprising because a thermal sheet with no pre-stored stress should display saddle-like shapes with zero overall mean curvature~\cite{Mizuochi2014}. It has been hypothesised that either the chirality of the linkages~\cite{Klotz2024} or redundant linking at the periphery~\cite{He2023} may be the reason behind the overall shape of the network. 

In our previous work~\cite{Ramakrishnan2024}, we found qualitative evidence that some maxicircles were located at the periphery of the network. In the previous section, we have quantitatively demonstrated that indeed maxicircles display a preference to be located at the periphery kDNAs. In this section we now ask if the peripheral positioning of the maxicircles may affect the overall curvature of the network. 

To do this we performed molecular dynamics simulations in LAMMPS~\cite{Plimpton1995a} of model kDNA networks with and without maxicircles (see Methods and SI). These networks are created using planar graphs and the DNA circles are modelled as coarse-grained ring polymers made by beads connected by FENE springs~\cite{Kremer1990}. The minicircles are interlinked to each other with random chirality and valence 3~\cite{Chen1995}. 
The maxicircles are randomly intertwined within a planar minicircles network compressed to resemble the disk-shape assumed by kDNA in vivo. This is done under two different hypotheses: i) Maxicircles are found at the border (LB networks) or ii) Maxicircles are found within the kDNA disk (LD networks). In both cases the networks are constructed by first confining the position of maxicircles within the kDNA through external potentials (later removed), and then slowly introducing a steric interaction between minicircles and maxicircles. This, together with FENE bonds in between bonded beads, ensures that the topolgy of the kDNA network is conserved~\cite{Kremer1990}.
After initialisation, we remove all constraints and equilibrate the system to relax any stress introduced during the preparation. We then perform production runs for at least $1.5\times 10^9$ timesteps. To verify that the system is truly in equilibrium we checked that $95\%$ of the values of the kDNA radius of gyration are within two standard deviations from the mean, i.e. that the size of the kDNA is not evolving or drifting during the production run. 

From our simulations we computed the Gaussian and mean curvatures of the networks (see Methods) and obtained the results in Fig.~\ref{fig:simbuckling}b. Configurations consisting of only minicircles display zero mean curvature compatible with a saddle shape. This is consistent with previous simulations of flat membranes~\cite{Mizuochi2014} and kDNA models~\cite{Luengo-Marquez2024,He2023}. 
Adding maxicircles in random locations inside the disc reduces the transition probability between the two states (positive and negative Gaussian curvatures) but the system is still able to switch from one state to the other and the total mean curvature remains close to zero. Interestingly, only when we placed the maxicircles along the border we observed net positive mean curvatures compatible with the experimentally-observed buckled shapes~\cite{Klotz2020}. We argue that this is because they constrain the total perimeter of the disc, acting as an elastic band running along the border of the kDNA~\cite{He2023} (see SI). Arguably, trypanosome species that have a different network replication and organisation, like \emph{T. Brucei}, may retain the a saddle-like shape in solution. We hope to test this hypothesis in the near future.

\subsection*{Dynamics of maxicircles sequences within the kDNA}

The main advantage of labelling DNA circles sequences with dCas9-QDs is that we can quantify their dynamics in real time. We employed a fast spinning-disk confocal to record 2-color movies at $>$10 fps of labelled kDNAs and QDs diffusing in a glycerol solution, optimised to record the dynamics of the kDNA (see Methods, and Fig.~\ref{fig:dynamics}a). 


The QD signal was processed with a tracking algorithm (see Methods) which allowed us to obtain the 2D location of each QD over time (see fig.~\ref{fig:dynamics}b). To remove the global translation of the kDNA, we also tracked the position of the kDNA COM using the DNA signal. We then computed the mean squared displacement of the single QDs in the frame of reference of the kDNA COM, i.e. 
$$g_2(t) = \langle [(\bm{r}_i(t+t_0) - \bm{r}_{CM}(t+t_0)) - (\bm{r}_i(t_0) - \bm{r}_{CM}(t_0)) ]^2 \rangle \, .$$ 

As one can appreciate from Fig.~\ref{fig:dynamics}d), $g2(t)$ displays two subdiffusive regimes, until it reaches a long time plateau at $t>10$ s. The subdiffusive, and hence correlated, motion of QDs bound to the maxicircles is to be expected, as the topological links create a correlation similar to that of bonded segments in tethered membranes~\cite{Mizuochi2014}. However, in classic Rouse dynamics of single polymers one expects $g2$ to plateau around the size of the polymer $R_g^2$. In this case instead, we observe a slower-than-Rouse dynamics~\cite{Halverson2011dynamics} ($g_2 \sim t^{\alpha}$, $\alpha < 0.5$) and a transition to the plateau when QDs have, on average, diffused less than $0.5$ $\mu$m, a distance far smaller than the size of the kDNA (around 4 $\mu$m). This points to a slower dynamics than that of classic 1D polymer segments. It also suggests a dynamics possibly slower of that of tethered membranes~\cite{Mizuochi2014}.

Additionally, we used our time-resolved imaging to infer the elastic properties of the network in a manner akin to that done for polymeric networks~\cite{Sorichetti2023}. We can interpret $g_2(t)$ as reflecting the fluctuations of DNA maxicircles sequences within a tethered, or crosslinked, structure. According to the equipartition theorem, at large times we expect $g_2(t)$ to be proportional to the thermal energy and inversely proportional to the effective kDNA stiffness i.e. $ g_2(t \to \infty) = 3k_BT/\kappa$. In Fig.~\ref{fig:dynamics}d, one can appreciate that the extent of these fluctuations are limited at around 0.2 $\mu$m$^2$, and we thus obtain $$ \kappa = \dfrac{3 k_B T}{0.2 \mu m^2} = 0.06 \dfrac{pN}{\mu m} \, ,$$ which is in line (although slightly smaller than) with our previous estimations based on AFM images of 0.1 pN/$\mu$m~\cite{He2023,Ramakrishnan2024}. We hypothesise that by removing (or linearising) some of the mini-circle classes forming the kDNA, we should expect a smaller effective stiffness. 

Another way to analyze the dynamics of the QDs that has the benefit of naturally removing the roto-translational motion of the kDNA is to consider the distance vector between pairs of QDs as $\bm{d}_{ij} (t)= \bm{r}_i(t) - \bm{r}_j(t)$ (see Fig.~\ref{fig:dynamics}a,c). We then compute its MSD as 
$$ \textrm{dMSD}(t) = \langle \left[ \bm{d}_{ij}(t-t_0) - \bm{d}_{ij}(t_0)\right]^2 \rangle \, ,$$ 
where the average is performed over initial times $t_0$ and over pairs of QDs. The MSD of the distance vector, dMSD, is a quantity typically measured when tracking genomic sites \emph{in vivo}~\cite{Gabriele2022, Yesbolatova2022} and could therefore be compared to experimental values of genome dynamics within the cell. As one can appreciate in Fig.~\ref{fig:dynamics}e, dMSD displays a strong subdiffusive regime, where $\textrm{dMSD(t)} \sim t^{0.25}$, somewhat slower than the exponent expected for tethered membranes~\cite{Mizuochi2014}. 

The dynamics of the distance between QDs can also yield information on the network stiffness. Indeed, by plotting all the QDs distance traces as a function of lag-time, i.e. $d(t)=|\vec{d}(t-t_0)|$, and normalised by the mean distance $\langle d \rangle$ we observe that they are all contained within $1 \pm 0.2$ (Fig.~\ref{fig:dynamics}f). More precisely, we obtain an average mean distance between QDs of $\bar{\langle b \rangle} = 2.16$ $\mu$m and an average standard deviation of $\sigma_{\langle d \rangle} = 0.17$ $\mu$m (see Figs.~\ref{fig:dynamics}g,h). Again from the equipartition theorem we expect that the dynamics of the QDs' distance should be equivalent to that of points connected by an effective spring with stiffness $3 k_B T/ \sigma_{\langle d \rangle}^2 = 0.43$ pN/$\mu$m, which also in excellent agreement with our previous estimate~\cite{He2023}. We argue that measuring the dynamics and the effective stiffness experienced by different DNA sequences, we will be able to infer inhomogeneous structures within the kDNA network. In turn, we could potentially extend this analysis to other complex genomes, therefore mapping their elastic properties \textit{in situ}. 

In summary, this is the first time we could obtain direct measurements on the stiffness of kDNA networks using time-resolved imaging and have found that these measurements are in broad agreement with the estimations from AFM images (which were only based on the network structure). We argue that our dynamic data could be directly compared with simulations of tethered and topologically interlocked membranes to better understand the dynamic scaling of Olympic-like networks compared with traditional crosslinked ones. 

\section{Discussion}
 
Understanding the spatial organisation of complex genomes is a question that is fascinating and ubiquitous. We have here shown that QD-labelled dCas9 can be used as ``beacons'' to map the location and dynamics of DNA sequences in a topologically complex genome such as the kDNA (Fig.~\ref{fig:labelling}). 

We used this method to map the location of kDNA maxicircles, which have previously been hypothesised to populate the outer part of kDNA networks~\cite{nabelshnur,Ramakrishnan2024}. Indeed, using our method we have quantitatively demonstrated that maxicircles are preferentially located at the periphery (Fig.~\ref{fig:distrmaxi}). On the contrary, we showed (see SI) that major and minor minicircle classes are mostly uniformly distributed over the kDNA network. Though there is no solid experimental evidence for how the replicated mini- and maxi-circles are redistributed within the kDNA during replication, we argue that the positioning of maxicircles at the periphery may play a structural function, for instance in establishing correct nabelschnur structure~\cite{nabelshnur} and in the partitioning of the kDNA to the daughter cells.  

We then asked if the peripheral location (and interlinking~\cite{Ramakrishnan2024}) of maxicircles played a role in determining the buckled shape of kDNA in solution~\cite{Klotz2020}. To answer this question we performed MD simulations of different kDNA topologies, including without maxicircles, or with maxicircles linked throughout the kDNA or only at the periphery. We observed that in the latter case the network buckles and displays the largest mean curvature (Fig.~\ref{fig:simbuckling}). We therefore argue that the observed buckling of kDNA in solution is not due to the chiral arrangement of the minicircle links, but it is instead due to the positioning of the maxicircles.   

Finally, we used our dCas9 labelling technique to track the dynamics of maxicircle sequences (Fig.~\ref{fig:dynamics}). We discovered a largely subdiffusive dynamics, slower than the dynamics seen in simulations of tethered membranes~\cite{Mizuochi2014}. By measuring the fluctuations of the dCas9 proteins with respect to either the centre-of-mass (COM) of the kDNA or with respect to each other, we obtained direct measurements of the kDNA network effective stiffness finding values $\kappa \simeq$ 0.06 - 0.4 pN/$\mu$m which are in excellent agreement with our previous estimate of 0.1 pN/$\mu$m based on AFM images alone~\cite{He2023,Ramakrishnan2024}. This measurement confirms a previous hypothesis that the kDNA is an ``ultra-soft'' 2D polymeric membrane, especially when compared with lipid bilayers or other 2D structures which typically display stiffnesses $\simeq 1 \mu N/\mu m$, i.e. $10^6$ times larger. 

We expect that our method could be applied to mapping the location and dynamics of DNA sequences in other complex genomes, for instance in ``genome-in-a-box'' set ups~\cite{Birnie2021} or even DNA origami, and could in turn provide information on the material properties of these structures.   

\section*{Acknowledgements}
DM thanks the Royal Society for support through a University Research Fellowship. This project has received funding from the European Research Council (ERC) under the European Union's Horizon 2020 research and innovation program (grant agreement No 947918, TAP). This work was supported by funding for the Wellcome Discovery Research Platform for Hidden Cell Biology [226791] and we gratefully acknowledge support from the Microscopy core.

\bibliography{library1,library}
\end{document}